\documentclass[showpacs,pra,aps,superscriptaddress,twocolumn,floatfix]{revtex4}
\usepackage{graphicx,float,amsmath,amssymb}
\usepackage{psfrag}
\newfont{\ensmathquatorze}{msbm10 scaled 1400}
\newfont{\ensmathonze}{msbm10 scaled 1100}
\newfont{\ensmathdix}{msbm10}
\newfont{\ensmathneuf}{msbm10 scaled 833}
\newfont{\ensmathhuit}{msbm10 scaled 694}
\newfam\ensmathfam
\textfont\ensmathfam=\ensmathonze
\scriptfont\ensmathfam=\ensmathdix
\scriptscriptfont\ensmathfam=\ensmathhuit

\renewcommand{\leq}{\leqslant}

\newcommand{\xho}{x_{\scriptscriptstyle {\rm HO}}}
\newcommand{\pho}{p_{\scriptscriptstyle {\rm HO}}}

\begin{document}

\title{Wavepacket Dynamics in Nonlinear Schr\"odinger Equations}

\author{S. Moulieras}
\affiliation{Laboratoire de Physique Th\'eorique et Mod\`eles Statistiques, CNRS, Universit\'e Paris Sud, UMR8626, 91405 Orsay Cedex, France}
\author{A. G. Monastra}
\affiliation{Gerencia Investigaci\'on y Aplicaciones, Comisi\'on Nacional de Energ\'\i a At\'omica, Avda. General Paz 1499, (1650) San Mart\'\i n, Argentina}
\affiliation{Consejo Nacional de Investigaciones Cient\'\i ficas y T\'ecnicas, Avda. Rivadavia 1917, (1033) Buenos Aires, Argentina}
\author{M. Saraceno}
\affiliation{Gerencia Investigaci\'on y Aplicaciones, Comisi\'on Nacional de Energ\'\i a At\'omica, Avda. General Paz 1499, (1650) San Mart\'\i n, Argentina}
\author{P. Leboeuf}
\affiliation{Laboratoire de Physique Th\'eorique et Mod\`eles Statistiques, CNRS, Universit\'e Paris Sud, UMR8626, 91405 Orsay Cedex, France}


\begin{abstract}
Coherent states play an important role in quantum mechanics because of their unique properties under time evolution. Here we explore this concept for one-dimensional repulsive nonlinear Schr\"odinger equations, which describe weakly interacting Bose-Einstein condensates or light propagation in a nonlinear medium. It is shown that the dynamics of phase-space translations of the ground state of a harmonic potential is quite simple: the center follows a classical trajectory whereas its shape does not vary in time. The parabolic potential is the only one that satisfies this property. We study the time evolution of these nonlinear coherent states under perturbations of their shape, or of the confining potential. A rich variety of effects emerges. In particular, in the presence of anharmonicities, we observe that  the packet splits into two distinct components. A fraction of the condensate is  transferred towards uncoherent high-energy modes, while the amplitude of oscillation of the remaining coherent component is damped towards the bottom of the well.
\end{abstract}

\pacs {05.60.Gg;67.85.De;42.50.Md}

\maketitle

\section{Introduction}

Coherent states were introduced in quantum mechanics by Schr\"odinger in 1926 to describe minimum uncertainty wave packets that satisfy the correspondence principle. The standard coherent states are defined as translations of the Gaussian ground state of the harmonic oscillator potential. The peculiarity of those states is that, during the time evolution in such a potential, they remain of minimum uncertainty at all times. This remarkable quasi-classical evolution is highly non trivial in quantum mechanics, the general rule being the spreading of the wave packet and the delocalization of the probability density. The harmonic oscillator coherent states arise in systems whose dynamical symmetry group is the Heisenberg-Weyl group. They can be generalized to systems with different symmetry groups, like the SU(2) spin coherent states, and appear in a wide range of physical situations \cite{cs,gaz}.

If an initial Gaussian wave packet is subjected to the action of an anharmonic potential, it will generally spread out. In some cases, after the initial spreading, the quantum state may, periodically, come back almost completely to its initial state. This revival of the wave packet occurs in systems where the spectrum may be expanded locally in terms of a quantum number, a characteristic situation of one-dimensional integrable Hamiltonian systems \cite{revival,gaz}. In contrast, if the corresponding classical dynamics is chaotic, the wave packet will spread and relax towards the phase space chaotic region, with time dependent fluctuations of the density that reflect interference effects. The structure of the underlying classical Hamiltonian thus has a strong influence on the dynamics of the packet, and may produce quite different effects depending on the integrable or chaotic nature of the classical dynamics \cite{chdyn}.

Here we are interested in a situation where the classical dynamics is simple, we consider integrable one-dimensional Hamiltonian systems. However, the difficulty is related to the more general character of the quantum dynamics considered, since we include nonlinear terms in the Schr\"odinger equation. The resulting nonlinear Schr\"odinger equation (the Gross-Pitaevskii equation, GPE) has a wide range of physical applications. It emerges, in particular, in two important cases: in the description of a Bose-Einstein condensate (BEC) of weakly interacting particles \cite{stri_pita}, and in the description of electromagnetic waves (light) propagating through a nonlinear medium \cite{boyd}.

The first point we are interested in is to determine if, in the nonlinear case, there still exist coherent states, in the sense of a set of initial states that are able to propagate in time without spreading or changing their shape. This question is particularly relevant in the context of BECs, since the mere existence of a coherent motion means, physically, that the condensate is preserved in time and that the atoms do not diffuse to different modes during the motion.  We consider here the particular case of a positive nonlinear coefficient, which corresponds to a BEC of repulsive interactions, or to a defocusing medium in nonlinear optics. The most elementary expectation would be that the additional repulsive nonlinear term in the Schr\"odinger equation enhances the spreading of an initial wave packet. This is of course true for the free propagation. However, as in the case of the linear Schr\"odinger equation, we find that a particular role is played by the harmonic confining potential. For that potential it is shown that the phase-space translations of the nonlinear ground state behave as coherent states, e.g. during the time evolution the center of the packet follows a classical phase-space trajectory, without any change of its shape. These translations therefore constitute a set of nonlinear coherent states which will be properly defined in section~\ref{2a}. This behavior is specific of the harmonic potential. Furthermore, we study the stability of the nonlinear coherent states under deformations of their shape. For small deformations, the packet remains coherent and its center follows the corresponding classical trajectory, with superimposed small shape oscillations of frequency given by the multipole modes of the ground state. Remarkably, this result holds also for large initial perturbations. For instance, the motion of a very compressed initial Gaussian state can be decomposed into a standard dipolar motion of its center and a superimposed large amplitude shape expansion and compression cycle.

The next relevant question concerns the evolution of a nonlinear coherent state subjected to an arbitrary 1D confining potential. In contrast to the linear case, when both anharmonicities and nonlinearities are present the spreading and revival of the packet are not observed, and a new phenomenology emerges.We find, as in previous studies \cite{li}, that for small anharmonicities and small amplitudes of oscillation the packet keeps, to a good approximation, its coherence. Its center follows a classical trajectory with superimposed small shape fluctuations. However, as the anharmonicity or the amplitude increase, a new process appears. The packet splits into two components, where part of the packet is damped towards the bottom of the potential, while the rest leaves the packet to form an uncoherent higher energy phase-space cloud.

\section{Wavepacket dynamics in a harmonic potential}
\subsection{Coherent states of the Gross-Pitaevskii equation}\label{2a}

We consider the one dimensional time-dependent Gross-Pitaevskii equation:
\begin{equation}\label{EGP}
\begin{split}
i \hbar \frac{\partial \Psi (x,t)}{\partial t} = &-\frac{\hbar ^2}{2 m} \frac{\partial^2 \Psi(x,t)}{\partial x^2} + V(x) \Psi(x,t) \\
& + \left( g N |\Psi(x,t)|^2 - \mu \right) \Psi(x,t) \ ,
\end{split}
\end{equation}
which describes, in the mean-field approximation, the dynamics of a Bose-Einstein condensate of $N$ identical bosons, in the presence of repulsive interactions ($g>0$), in an external potential $V(x)$ \cite{stri_pita}. Here, $\Psi(x,t)$ is the normalized wavefunction of the condensate, $m$ the mass of each particle, $g$ the interaction constant, and $\mu$ the chemical potential. Aside from cold atom physics, it has been shown that Eq.~(\ref{EGP}) provides an accurate description of many interesting physical problems, among which we can mention hydrodynamics \cite{zakharov}, or nonlinear optics \cite{boyd}. In the latter case, the 1D GPE can be derived from the propagation of light in a two-dimensional nonlinear medium, under both the monochromatic and the paraxial approximations.

We assume that $V (|x|\rightarrow\infty)\rightarrow\infty$, and look for solutions of the GPE which evolve in time without changing their shape. We thus seek for solutions in the form
\begin{equation}\label{expression}
\Psi(x,t)= \phi(x-x_0(t),t) \exp{\left(\frac{i  p_0 (t)}{\hbar} \left(x-\frac{x_0 (t)}{2}\right) \right)} \ ,
\end{equation}
where $x_0(t)$ and $p_0(t)$ are real functions of time. This solution represents a time dependent evolution in which the wavefunction is translated along the phase-space trajectory $(x_0(t),p_0(t))$. The substitution of
Eq.(\ref{expression}) into Eq.(\ref{EGP}) gives
\begin{equation}\label{inter}
\begin{split}
i \hbar \left. \frac{\partial \phi}{\partial t}\right|_{x-x_0(t),t} &= \frac{-\hbar ^2}{2 m}\left. \frac{\partial^2 \phi}{\partial x^2} \right|_{x-x_0(t),t} + V(x) \phi(x-x_0(t),t) \\ & + \left( g N |\phi(x-x_0(t),t)|^2 - \mu \right) \phi(x-x_0(t),t) \\ &+ i \hbar \left( \dot{x_0}(t) - \frac{p_0(t)}{m} \right) \left. \frac{\partial \phi}{\partial x}\right|_{x-x_0(t),t} \\ &-\frac{p_0(t)}{2}\left( \dot{x_0}(t) - \frac{p_0(t)}{m} \right) \phi(x-x_0(t),t) \\ & + \dot{p_0}(t) \left(  x - \frac{x_0(t)}{2} \right) \phi(x-x_0(t),t) \ ,
\end{split}
\end{equation}
where $\dot{x_0}(t) \equiv d x_0 / dt$ and $\dot{p_0}(t) \equiv d p_0 /dt$. Equation (\ref{inter}) takes a simpler form if the phase space trajectory $(x_0(t),p_0(t))$ coincides with a trajectory of the corresponding classical non-interacting problem
\begin{eqnarray*}
\dot{x_0}(t) &=& \frac{p_0(t)}{m}\\
\dot{p_0}(t) &=& -\left. \frac{\partial V}{\partial x}\right|_{x_0(t)} \ .
\end{eqnarray*}
Making the change of notation $ x-x_0(t) \to x$, Eq.~(\ref{inter}) simplifies to
\begin{equation}\label{general}
\begin{split}
i \hbar \frac{\partial \phi}{\partial t} &= -\frac{\hbar ^2}{2 m}\frac{\partial^2 \phi}{\partial x^2} + \left( g N |\phi|^2 - \mu \right) \phi\\ & +  \left(V(x+x_0(t)) -\left. \frac{\partial V}{\partial x}\right|_{x_0(t)} \left( x + \frac{x_0(t)}{2} \right)  \right) \phi \ ,
\end{split}
\end{equation}
in which $\phi$ and its derivatives are now evaluated in $(x,t)$. Equation (\ref{general}) shows that, in the reference frame of the classical trajectory, the particle feels a time-dependent potential. In the new reference frame, the coherent state should be a stationary state of Eq.~(\ref{general}). The stationarity condition imposes a time-independent potential. This leads, for any $x$, to the condition
\begin{equation}\label{indep}
\frac{d}{d x_0} \left[ V(x+x_0) - \left. \frac{\partial V}{\partial x}\right|_{x_0} \left( x + \frac{x_0}{2} \right) \right] =0 \ .
\end{equation}
In particular, for $x=0$ it takes the form
\begin{equation}\label{indep2}
 x_0 \frac{\partial^2 V}{\partial x_0^2} - \frac{\partial V}{\partial x_0} =0 \ .
\end{equation}
This equation is satisfied if and only if $V(x)$ is a quadratic function of $x$. Hence, the only function that produces, in the new reference frame, a time-independent potential is the harmonic one. Finally, for a harmonic potential and in the reference frame that follows the classical phase-space trajectory, the quantum equation of motion takes the form
\begin{equation}\label{harm}
i \hbar \frac{\partial \phi}{\partial t} = -\frac{\hbar ^2}{2 m}\frac{\partial^2 \phi}{\partial x^2} + V(x)\phi + \left( g N |\phi|^2 - \mu \right) \phi \ .
\end{equation}
Therefore, the coherent states of the GPE are defined by its stationary states, that satisfy the equation
\begin{equation}\label{phi0}
 -\frac{\hbar ^2}{2 m}\frac{\partial^2 \phi_0}{\partial x^2} + V(x)\phi_0 + g N |\phi_0|^2 = \mu  \phi_0 \ .
\end{equation}
It follows that, for a harmonic potential,
\begin{equation}\label{sol}
\Psi_0(x,t)= \phi_0(x-x_0(t)) \exp{\left(\frac{i  p_0 (t)}{\hbar} \left(x-\frac{x_0 (t)}{2}\right) \right)} \ ,
\end{equation}
is a time-dependent exact solution of Eq.~(\ref{EGP}). Here, $(x_0(t),p_0(t))$ is a phase-space trajectory of the corresponding non interacting classical system. In other words, the time evolution of the wave packet defined by Eq.~(\ref{sol}) reduces simply to the time evolution of its center, that follows a classical trajectory. Among the different possible stationary states $\phi_0$ of Eq.~(\ref{phi0}), it is customary to define as the standard coherent state the ground state, which minimizes the energy as well as its spatial extension \cite{com1}. From now on we refer to the set $\Psi_0(x,t)$, with $\phi_0$ defined as the ground-state of Eq.~(\ref{phi0}) and $x_0(0)$ and $p_0(0)$ arbitrary, as the set of \textit{nonlinear coherent states}.

It is easy to see that the previous results are not only valid for a quadratic nonlinearity of the GPE, but that they hold in fact for an arbitrary exponent $ \sim g N |\Psi(x,t)|^\alpha$. This remark extends our results to a large family of nonlinear Schr\"odinger equations.

\begin{figure}
\includegraphics*[width=1.0\linewidth,angle=0]{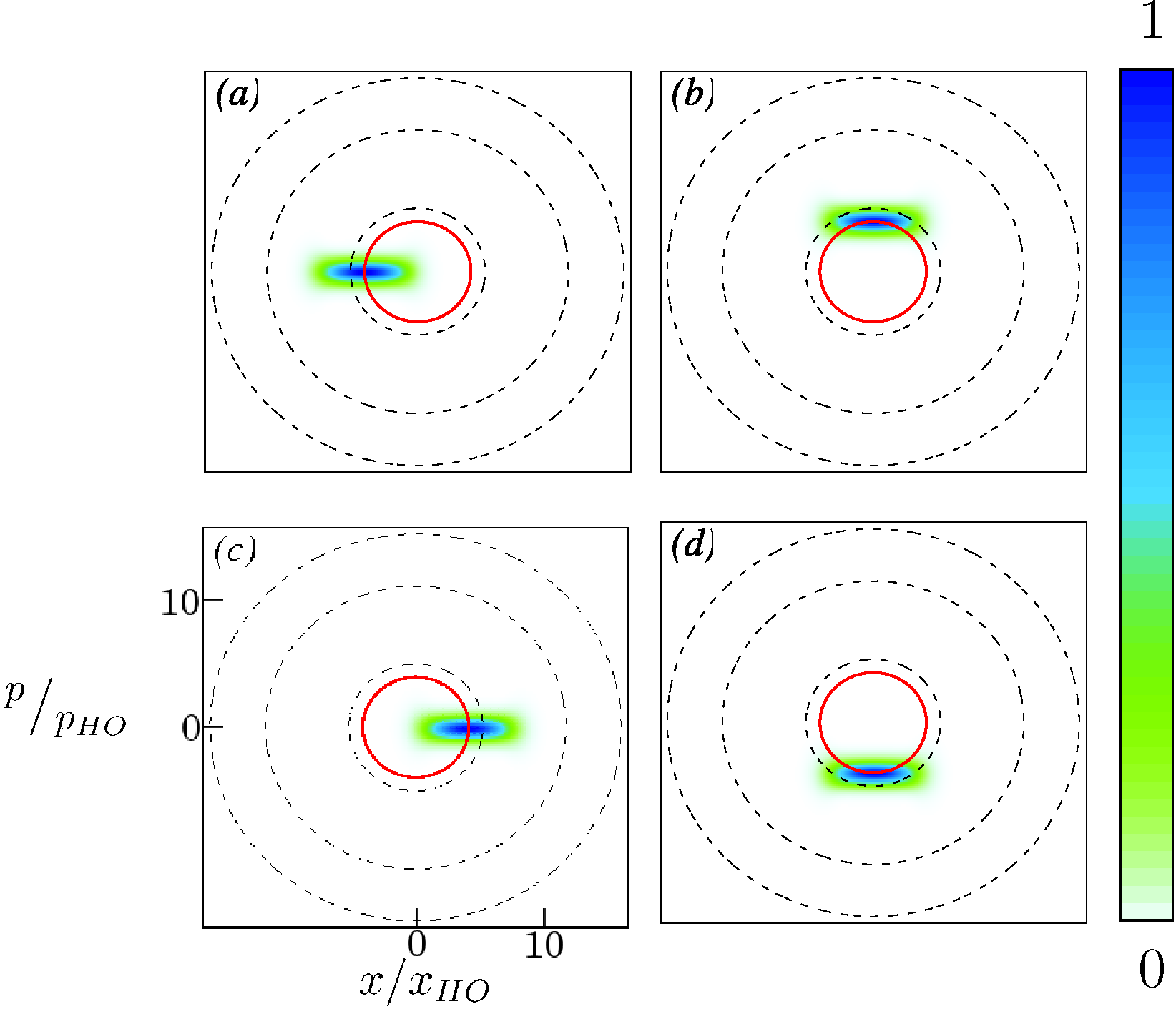}
\caption{(color online) Time evolution of a shifted ground state of the GPE with a harmonic confining potential, with parameters $\gamma=115$, $d=5$.  Husimi representations of the wavefunction are given at times $\omega t/ (2\pi)=0$ (a), 0.25 (b), 0.5 (c), 0.75 (d). The (red) full curve is the classical trajectory of the corresponding linear problem of energy given by the center of the initial packet.}\label{fig1}
\end{figure}

In order to illustrate the previous results, we have numerically computed the time evolution of Eq.~(\ref{EGP}), and plotted the phase-space Husimi distribution of the wavefunction at different times. This distribution is defined as
$$
{\cal H} (x,p,t) = |\displaystyle \langle x \ p | \Psi (t) \rangle|^2 \ ,
$$
where $|x \ p \rangle$ is a standard linear harmonic oscillator coherent state centered around the phase-space point $(x,p)$, whose $x$ representation reads:
\begin{equation}
\begin{split}
\displaystyle \langle x | x_0 \ p_0 \rangle = &\left( \frac{m \omega}{\pi \hbar}\right) ^{1/4} \exp{\left(-\frac{ (x- x_0)^2 }{\xho^2}\right)} \\
& \times \exp{\left(\frac{i  p_0 }{\hbar} \left(x-\frac{x_0}{2}\right) \right)} \ .
\end{split}
\end{equation}
The typical width of a standard coherent state in the $x$ and $p$ directions is $\xho\equiv (2 \hbar / m \omega)^{1/2}$ and  $p_\text{HO}\equiv \left(2 \hbar m \omega\right)^{1/2}$, respectively. To obtain Fig.~(\ref{fig1}), we numerically calculate the ground state $\phi_0 (x)$ of the Gross-Pitaevskii equation in a harmonic trap, $V(x)= \frac{1}{2}m\omega^2 x^2$, and then compute the time evolution of a translated ground state, $\Psi(x,t=0) = \phi_0 (x+d)$. In order to characterize the intensity of the nonlinearity, it is convenient to define a dimensionless parameter. In terms of the characteristic width $\xho$ and energy $\hbar \omega$ of the ground state of the noninteracting harmonic oscillator, we define the parameter $\gamma=2 g N/ (\xho \hbar \omega)$,
\begin{equation}
\gamma \equiv \sqrt{\frac{2 m}{\hbar \omega}} \frac{g N}{\hbar} \ .
\end{equation}

As predicted above, in the nonlinear case the wave packet dynamics reduces to a simple phase-space translation of its center, that follows the corresponding classical trajectory (full (red) curve in the figure). During this process, its shape does not vary in time, and there is no rotation either. In particular, the shape of its projection onto the $x$ axis does not change in time.

This behavior qualitatively differs from the dynamics of the linear Schr\"odinger equation (non-interacting case), where the motion of an arbitrary initial wavefunction in a harmonic trap consists in a phase-space rigid rotation with respect to the origin \cite{tannor}: defining $z= x /\xho + i p/p_\text{HO} $, it is known that the linear evolution of an arbitrary initial Husimi distribution $\mathcal{H}_0 (z)$ in a harmonic oscillator reads:
\begin{equation}
\mathcal{H} (z, t) \equiv \mathcal{H} (x, p, t) = \mathcal{H}_0 (z\, e^{i \,\omega \, t}) \, .
\end{equation}
This implies a rigid phase-space rotation of any initial state.

\begin{figure}
\includegraphics*[width=1.0\linewidth,angle=0]{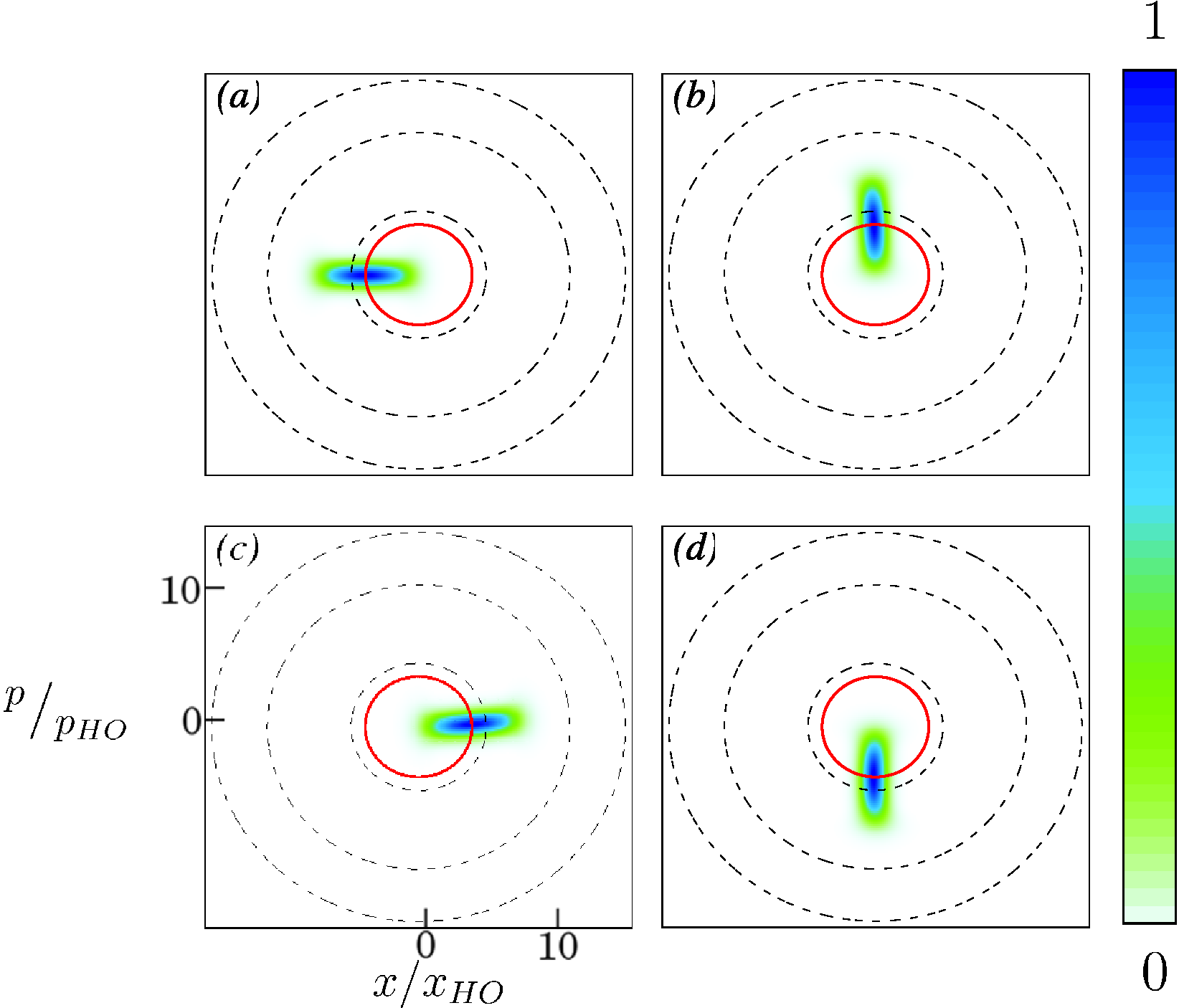}
\caption{(color online) Time evolution of the linear Schr\"odinger equation with a harmonic potential with parameters $\gamma=0$, $d=5$. The initial state is the same one as in Fig.~(\ref{fig1}) (a shifted groundstate of the GPE) .(a): Husimi representations of the wavefunction are given at times $\omega t/ (2\pi)=0$ (a), 0.25 (b), 0.5 (c), 0.75 (d).}\label{fig2}
\end{figure}

To stress the difference between the linear and the nonlinear dynamics, we plot in Fig.~(\ref{fig2}) the linear evolution of the same initial state as in Fig.~(\ref{fig1}). We observe that, in contrast to the nonlinear evolution, the initial packet now rotates as it follows the classical trajectory, and therefore changes its shape as a function of time in the position representation. The coherent state of the linear case corresponds, necessarily, to a perfectly spherical Gaussian initial packet, a shape which is invariant under rotations in any representation.

Remark that the classical trajectory followed by the center of the packet has no dependence on the interaction parameter $g$. It is a classical trajectory of the noninteracting problem, fixed by the initial position of the packet. In particular, the frequency of the oscillation is independent of the interaction, a result demonstrated by Kohn \cite{kohn} for the cyclotron frequency of interacting particles, that was later on generalized to interacting particles in a parabolic confining potential \cite{bjh}.

The experimental realization of nonlinear coherent states, as well as the control of their initial phase-space location, is a natural procedure in the context of cold atom physics. This is because cold atoms are usually trapped in parabolic magnetic potentials, and the corresponding BEC is thus a coherent state centered at the bottom of the potential. Phase-space translations of that state are easily implemented  by a sudden shift of the trap with respect to the condensate. The study of dipolar oscillations were among the first experimental tests of excited collective states \cite{stringari}. More recently, dipole excitations were used to test transport properties of BECs across an impurity \cite{imp,appl,hulet} or through disordered potentials \cite{dis,appl,hulet}. Dipole oscillations were also proposed as a test of the existence of a superfluid phase for light moving in a nonlinear medium \cite{bibi}.

The quantum dynamics in the presence of nonlinearities is thus particularly simple if the initial state is a coherent state. What happens to an arbitrary initial state? We will explore in detail this question in the following sections, which will be particularly relevant in the context of nonlinear optics since, in contrast to BECs, in optics gaussians are the natural transverse intensity profiles.

\subsection{Stability of the oscillations}

In this section, we study the stability under deformations of the initial wave packet $\Psi_0(x,t)$ (Eq.~(\ref{sol})) in the presence of a harmonic confining potential $V(x)=\frac{1}{2}m \omega^2 x^2$. For that purpose, we look for solutions of the GPE having the form (\ref{expression}) and were $\phi(x,t)= \phi_0(x) + \delta \phi(x,t)$. Actually, the problem of the stability of the time dependent solution $\Psi_0(x,t)$ is equivalent to the problem of stability of the stationary ground state of Eq.~(\ref{EGP}). The first order expansion in $\delta \phi$ of Eq.(\ref{harm}) leads to
\begin{equation}\label{pert}
\begin{split}
i \hbar \frac{\partial \delta \phi}{\partial t} = &-\frac{\hbar ^2}{2 m}\frac{\partial^2 \delta \phi}{\partial x^2} + V(x)\delta \phi - \mu \delta \phi\\&+  g N \left(2|\phi_0|^2 \delta \phi + \phi_0^2  \delta \phi^* \right) \ ,
\end{split}
\end{equation}
which, with its complex conjugate equation, form the so called Bogoliubov-de Gennes (BdG) system. Since $\phi_0$ is real, the BdG system reduces to
\begin{equation}\label{stabi}
i \hbar \frac{\partial }{\partial t} \left[ \begin{array}{c} \delta \phi \\ \delta \phi^* \end{array} \right] = M \left[ \begin{array}{c} \delta \phi \\ \delta \phi^* \end{array} \right] \ ,
\end{equation}
where $ M = \begin{bmatrix} \Lambda &  g N \phi_0^2 \\  - g N \phi_0^2 & -\Lambda \end{bmatrix}$ and $\Lambda =-\frac{\hbar ^2}{2 m}\frac{\partial^2}{\partial x^2} + V(x) + 2 g N |\phi_0|^2 - \mu$. The stability of the solution $\phi_0$ is given by the sign of the eigenvalues $\hbar \omega_n$ of $M$, which are the energies of the elementary excitations $\left[ \begin{array}{cc} u_n, v_n \end{array} \right]$, given by
\begin{equation}
\hbar \omega_n \left[ \begin{array}{c} u_n \\ v_n \end{array} \right] = M \left[ \begin{array}{cc} u_n\\ v_n \end{array} \right] \ .
\end{equation}
Our calculations are the 1D equivalent of the 2D work of Ref.~\cite{chin}, and we will not give the technical details here. For instance, in the strongly interacting limit (the so called Thomas-Fermi limit), the spectrum is given, for $n \in \mathbb{N}^*$, by
\begin{equation}\label{spectrum}
\frac{\omega_n}{\omega}=\sqrt{\frac{n(n+1)}{2}} \ .
\end{equation}
This result shows that the frequencies become, in that limit, independent of the nonlinearity and that the $n=1$ dipolar excitation is unchanged, $\omega_1=\omega$. All eigenvalues are real, a fact that ensures the dynamical stability of the coherent state under small deformations.

In the following  we use a different method to test the stability of the motion of coherent states under shape deformations. We use the Virial theorem for the GPE \cite{vlasov, lushnikov} and, applying a variational principle, recover the former results as well as some extensions of their regime of validity. The Virial Theorem states that, for a solution $\Psi(x,t)$ of Eq.(\ref{EGP}), the average spatial extension $<x^2>$ of $\Psi(x,t)$ verifies
\begin{equation}\label{VT}
\partial^2_t <x^2> = \frac{1}{m} \left[ 4 E_K + 2 E_{NL} -2 \left< x \frac{\partial V}{\partial x} \right>\right] \ ,
\end{equation}
where
\begin{eqnarray}\label{defs}
E_K  &\equiv& \int \frac{\hbar^2}{2m} |\partial_x \Psi(x,t)|^2  dx \ , \\
E_{NL} &\equiv& \frac{g}{2} \int  |\Psi(x,t)|^4  dx \ , \\
E_{P} &\equiv& \int  V(x)|\Psi(x,t)|^2  dx \ ,
\end{eqnarray}
and $\partial_x \equiv \frac{\partial }{\partial x}$, $\partial_t \equiv \frac{\partial }{\partial t}$, $< A(x) > \equiv \int A(x) |\Psi(x,t)|^2 dx $ for any function $A(x)$. This theorem has been used in particular to study the collapse dynamics of a BEC. It is important to mention that Eq.~(\ref{VT}) follows from the fact that $\Psi$ extremizes the Gross-Pitaevskii functional $E[\Psi] = E_K + E_{NL} +E_{P}$. The quantity $E = E_K + E_{NL} +E_{P}$ does not depend on time. In the particular case $V(x)= \frac{1}{2} m \omega^2 x^2$, the relation $\left< x \frac{\partial V}{\partial x} \right> = 2 E_P $ leads to
\begin{equation}\label{VTHP}
\partial^2_t <x^2> = \frac{1}{m} \left[ 4 E_K + 2 E_{NL} -4 E_P\right] \ .
\end{equation}
For instance, for the non-interacting case, $g=0$, $E_{NL}=0$, and thus $E = E_K + E_{P}$ is a constant determined by the initial condition. Then, Eq.(\ref{VTHP}) simplifies to
\begin{equation}\label{linearviriel}
\partial^2_t <x^2> = - 4 \omega^2 \left( <x^2> + \frac{E}{m \omega^2} \right) \ .
\end{equation}
This means that, for any initial wavefunction, the spatial extension of $\Psi(x,t)$ is an oscillatory function of time, with frequency $2\omega$, a fact clearly seen in Fig.(\ref{fig2}). Indeed, since, as we mentioned previously, the dynamics in the non-interacting (linear) case of a harmonic oscillator is simply a rigid rotation in phase space, it is clear that every half-period of the oscillator the spatial extension comes back to its initial value.

We now take into account the presence of interactions, and more particularly, we assume to be in the Thomas-Fermi limit $\gamma \to \infty$. The reason for this assumption is that in this case an explicit form of the ground state $\phi^{TF}_0(x)$ is known
\begin{equation}\label{TFprofile2}
\phi^{TF}_0(x) = \sqrt{\frac{\mu - \frac{1}{2}m \omega^2 x^2}{gN}}
\end{equation}
(for $x^2 \leq 2 \mu /m w^2$, whereas $\phi^{TF}_0(x) =0$ for $x^2 > 2 \mu /m w^2$). In order to solve Eq.~(\ref{VTHP}), we assume that the wavefunction is able, during its time evolution, to follow the classical trajectory as well as to vary its spatial extension, denoted $L$. For $ |x-x_0(t)|<L(t)$ we write it in the form
\begin{equation}\label{ansatz}
\begin{split}
\Psi_{L}(x,t) = &C(L(t)) \sqrt{1 - \frac{(x-x_0(t))^2}{L(t)^2}} \\
&\times \exp{\left(\frac{i  p_0 (t)}{\hbar} \left(x-\frac{x_0 (t)}{2}\right) \right)} ,
\end{split}
\end{equation}
and $\Psi_{L}(x,t)=0$ if $ |x-x_0(t)|>L(t)$. In the latter expression, $C(L) = \sqrt{3/4L}$ ensures the normalization of $\Psi_L (x,t)$ at any time. Let us substitute Eq.(\ref{ansatz}) into the virial theorem (\ref{VTHP}), in which all terms depend only on $L(t)$, $x_0(t)$ and $p_0(t)$, respectively noted $L$, $x_0$ and $p_0$ for a matter of readability, and their derivatives $\dot{x_0} \equiv \frac{\partial x_0(t)}{\partial t} $, $\dot{p_0} \equiv \frac{\partial p_0(t)}{\partial t}$, and $\dot{L} \equiv \frac{\partial L(t)}{\partial t}$:
\begin{equation}\label{VTHP2}
\begin{split}
\frac{2}{5}(L \ddot{L} + \dot{L}^2)+2(x_0 \ddot{x_0} + \dot{x_0}^2) =&\\
4\frac{p_0^2}{2m^2} -2 \omega^2 (x_0^2 + \frac{L^2}{5})+\frac{3gN}{5mL} \ .
\end{split}
\end{equation}
Using the classical equations of motion, all the terms containing information concerning the classical trajectory vanish, and we finally obtain
\begin{equation}\label{VTTF}
L \ddot{L} + \dot{L}^2 = - \omega^2 L^2 +\frac{3gN}{2mL} \  .
\end{equation}
The equilibrium solution of the latter differential equation is $L_{eq}=\left( \frac{3gN}{2 m \omega^2} \right)^{1/3}$ which coincides with the usual spatial extension of the Thomas-Fermi solution. Let us now consider small deviations with respect to its extension, and write $L(t)$ in Eq.(\ref{VTTF}) in the form $L(t)=L_{eq} + \delta L(t)$. Performing a first order expansion in $u(t)\equiv \delta L(t) / L_{eq} \ll 1$, we get
\begin{equation}\label{sqrt3}
\ddot{u} + 3 \omega^2 u=0 \ ,
\end{equation}
which describes a periodic oscillatory motion of the width of the wave packet of frequency $\sqrt{3}\omega$. This frequency corresponds to the $n=2$ quadrupole mode of the excitation spectrum of Eq.(\ref{spectrum}). To summarize, in the two limiting situations  $\gamma=0$ and $ \gamma \to \infty$ the quadrupole deformations of the time dependent coherent state are stable and the corresponding frequencies are $2\omega$ and $\sqrt{3}\omega$, respectively.

\begin{figure}
\includegraphics*[width=1.0\linewidth,angle=0]{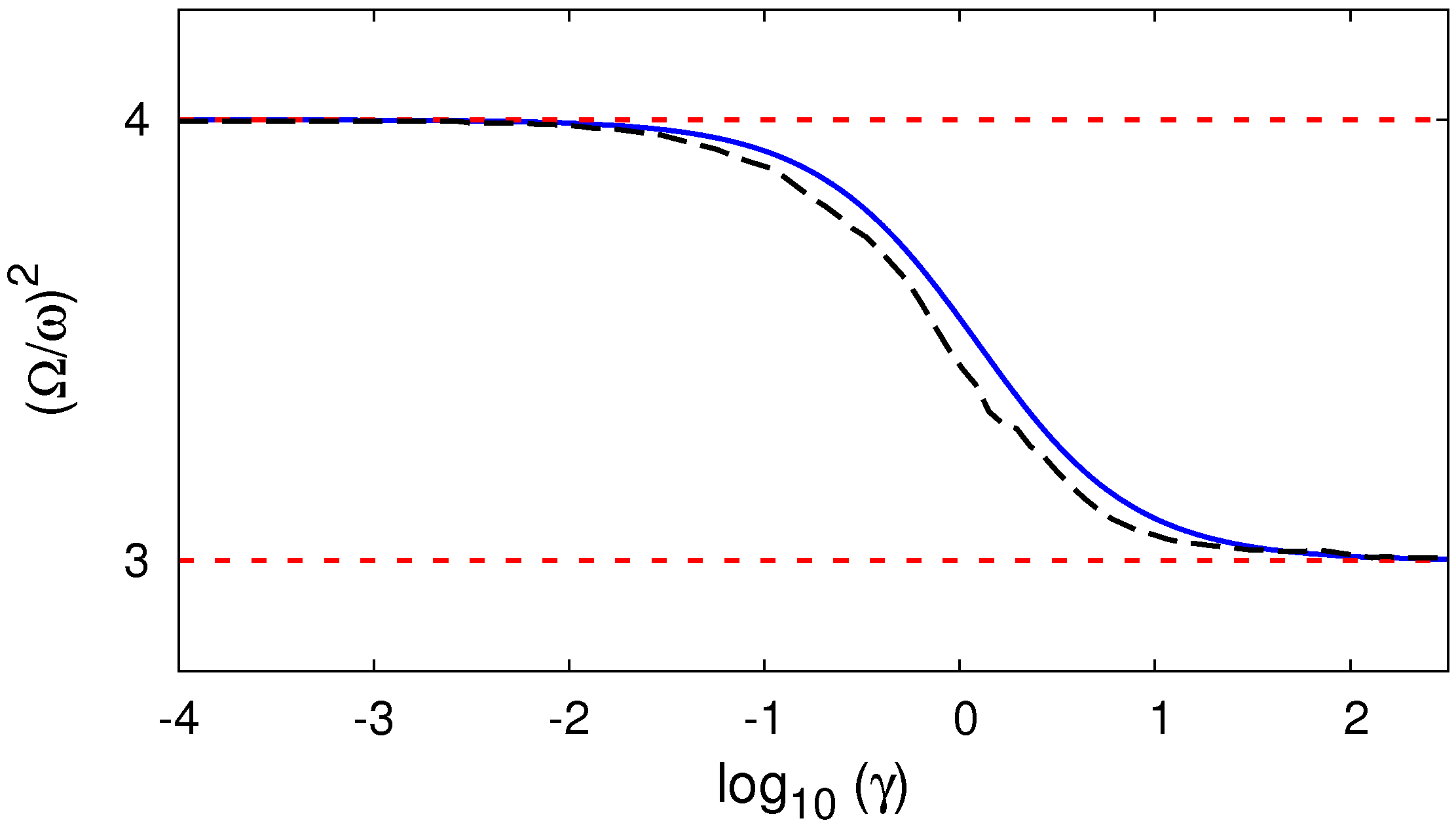}
\caption{(color online) Normalized square quadrupole frequency for different values of the nonlinear parameter $\gamma$. (Black) dashed line represents the numerically computed frequency, full (blue) line the analytical result obtained using a gaussian ansatz variational principle.}\label{fig4}
\end{figure}

In order to study the intermediate regime, for which we have no analytical expression of the ground state, we choose to use a normalized gaussian ansatz $\Phi^{\eta}(x,t)$ (which tends to the correct form in the absence of nonlinearities), with a time-dependent width $\eta(t)$
\begin{equation}\label{ansatzgauss}
\begin{split}
\Phi^{\eta}(x,t) = &\frac{1}{(2 \pi \eta^2(t))^{1/4}} \exp{\left(-\frac{ (x- x_0 (t))^2 }{4\eta(t)^2}\right)}\\
&\times \exp{\left(\frac{i  p_0 (t)}{\hbar} \left(x-\frac{x_0 (t)}{2}\right) \right)} \ .
\end{split}
\end{equation}
The same procedure as before leads to the following differential equation for $\eta(t)$
\begin{equation}\label{VTgauss}
\begin{split}
2(\eta \ddot{\eta} + \dot{\eta}^2)=\frac{\hbar^2}{2m^2\eta^2} -2 \omega^2 \eta^2+\frac{gN}{2\sqrt{\pi}m\eta} \ .
\end{split}
\end{equation}
Replacing in Eq.(\ref{ansatzgauss}) the stationary width $\eta_0\equiv \xho/2$ of the linear $g=0$ limit of Eq.(\ref{VTgauss}) gives the function $\Phi^{\eta_0}(x,t)$ which coincides with the well known definition of the usual coherent state of the harmonic oscillator, defined by the complex parameter $z= x_0 /\xho+ i p_0/\pho$. For a non zero interaction constant, $u(t)\equiv \eta(t) / \eta_0$ verifies
\begin{equation}\label{edgauss}
u\ddot{u}+\dot{u}^2 =\omega^2 \left[ \frac{1}{u^2} - u^2 + \frac{\gamma}{\sqrt{\pi}} \frac{1}{u} \right] \ .
\end{equation}
Let us denote $u_{eq}(\gamma)$ the strictly positive equilibrium solution of Eq.(\ref{edgauss}). $u_{eq}(\gamma)$ is an increasing function of $\gamma$, equal to $1$ for $\gamma=0$, and that tends to infinity in the limit $\gamma \rightarrow \infty$. Similarly as above, we perform a first order expansion writing $u(t)=u_{eq}(\gamma) + \delta u(t)$, and assuming $\delta u(t)\ll u_{eq}(\gamma)$, to obtain again a second order differential equation
\begin{equation}\label{quadgauss}
\ddot{\delta u} + \Omega^2 \delta u=0 \ ,
\end{equation}
where $\Omega$ is, in this approximation, the quadrupole frequency
\begin{equation}\label{quadgauss2}
\Omega^2 = \omega^2 \left(3 + \frac{1}{u_{eq}(\gamma)^4} \right) \ .
\end{equation}
Note that equations (\ref{edgauss}) and (\ref{quadgauss2}) have been already obtained by a variational principle in Ref.~\cite{li} including also the fourth order moment as time-dependent parameter. In the linear limit $\gamma=0$, $u_{eq}(0)=1$, and we recover $\Omega=2\omega$, as it should. In the other limit of strong nonlinearity, $u_{eq}(\gamma \rightarrow \infty)\rightarrow \infty$, and we recover $\Omega= \omega\sqrt{3}$, which is the correct result, as was shown previously. In Fig.~(\ref{fig4}) we plot the comparison of Eq.(\ref{quadgauss2}) for arbitrary $\gamma$ to a numerical calculation of the quadrupole frequency. Despite the fact that the gaussian ansatz is only correct in the linear limit, we see that it provides a quite good approximation of the quadrupole frequency for arbitrary $\gamma$. (The numerical simulation performed is the following: For any $\gamma$, we numerically compute the ground state of the GPE, we spatially shift it from the bottom of the potential, and apply a (norm-preserving) deformation (getting $\delta L(t=0)=0.1 L_{eq}$) in order to excite the quadrupole mode. Then, the real-time evolution of the GPE is computed, and the frequency which maximizes the Fourier transform on $\delta L(t)$ is finally found.)

\begin{figure*}
\includegraphics*[width=1.0\linewidth,angle=0]{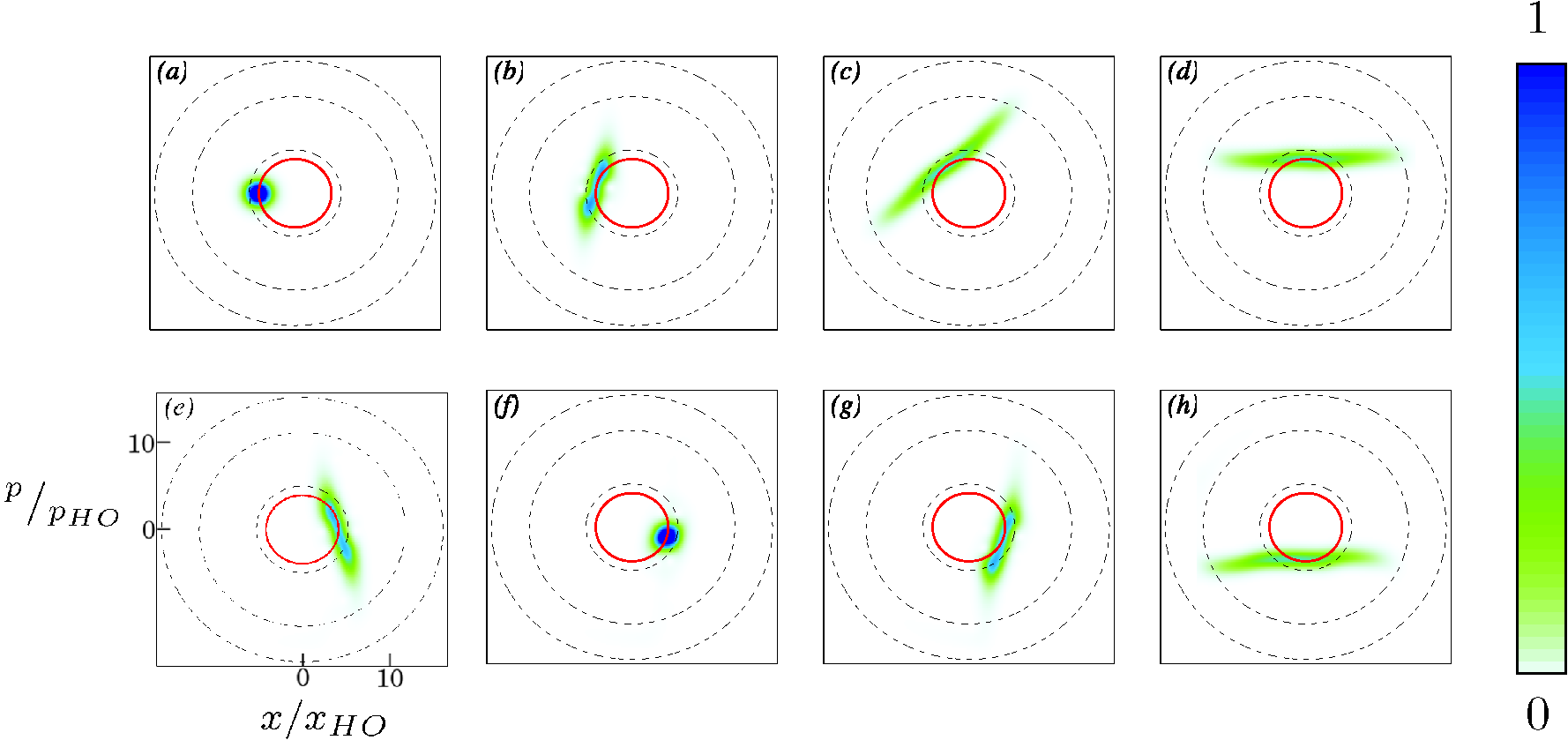}
\caption{(color online) Time evolution of the GPE with a harmonic confining potential with parameter $\gamma=115$. The initial state is the shifted Gaussian ground state of the linear problem. Husimi representations of the wavefunction are given at times $\omega t/ (2\pi)=0$ (a), 0.02 (b), 0.13 (c), 0.25 (d), 0.50 (e), 0.51 (f), 0.53 (g), 0.75 (h).}\label{fig3}
\end{figure*}

The previous results show the stability of the coherent states (and therefore of a condensate) under small shape perturbations when moving in a harmonic potential, and provide the typical frequencies involved. We have also numerically explored the evolution of packets whose initial shape strongly deviates from the coherent state. For instance, in Fig.~(\ref{fig3}) we show the nonlinear evolution of a Gaussian coherent state of the linear problem (defined as the translated Gaussian ground state of that problem). What is observed is the usual dipole oscillation following the corresponding classical trajectory with a superimposed large amplitude quadrupole vibration. The spatial width of the initial Gaussian state is small compared to the corresponding nonlinear state, see Fig.(\ref{fig1}). It follows that, because of the repulsive interactions, the packet strongly spreads in phase-space, predominantly in the $p$ direction (particles accelerate, see part (b) of the figure). This acceleration produces a spatial spreading of the packet, whose barycenter follows the corresponding classical trajectory (part (c)). At this point the expansion stops, compensated by the harmonic confinement, and a compression phase follows, to recover its initial shape. The process can start again. We have numerically computed the period of the expansion and compression cycle, and found a period (normalized to the harmonic oscillator period) $T_o/T \simeq 0.551$, which is close, but nevertheless different, from the quadrupole frequency predicted from Fig.~(\ref{fig4}) for the corresponding value of $\gamma$, $T_4/T\simeq 0.575$.

\section{Anharmonic external potential} \label{anharmonic}

We now explore the robustness of the motion of nonlinear coherent states when the considered potential differs from the harmonic oscillator. More generally, we wish to explore the nonlinear motion of initial wave packets under an arbitrary potential. Experimentally this is a relevant problem since anharmonic potentials are either use on purpose \cite{bssd}, or they come as corrections to the nearlly harmonic usual traps. From a theoretical point of view, frequency shifts and coupling of collective modes due to anharmonicities was explicitly investigated in the past \cite{li}. 

As an example we consider a potential of the form
\begin{equation}\label{nlpot}
V(x)=\frac{1}{2}m \omega^2 x^2 (1 + \alpha x^2) \ ,
\end{equation}
where $\alpha$ controls the strength of the anharmonicity. We consider as initial state the nonlinear coherent state of the corresponding harmonic oscillator, i.e. we compute the ground state of the nonlinear equation with $\alpha=0$ (the use of the true ground state does not qualitatively modify the results). This state is then shifted along the $x$ direction in order to locate the center of the packet at $x=-d$, with $d$ positive. The time evolution of such state is then computed for the full potential including the quartic term. As $d$ increases, the strength of the quartic term of the potential compared to the harmonic one increases. This strength is measured by the dimensionless parameter $\beta = \alpha d^2$. We thus study how the dynamics of the initial packet changes as a function of $\beta$.

\begin{figure*}
\includegraphics[width=1.0\linewidth,angle=0]{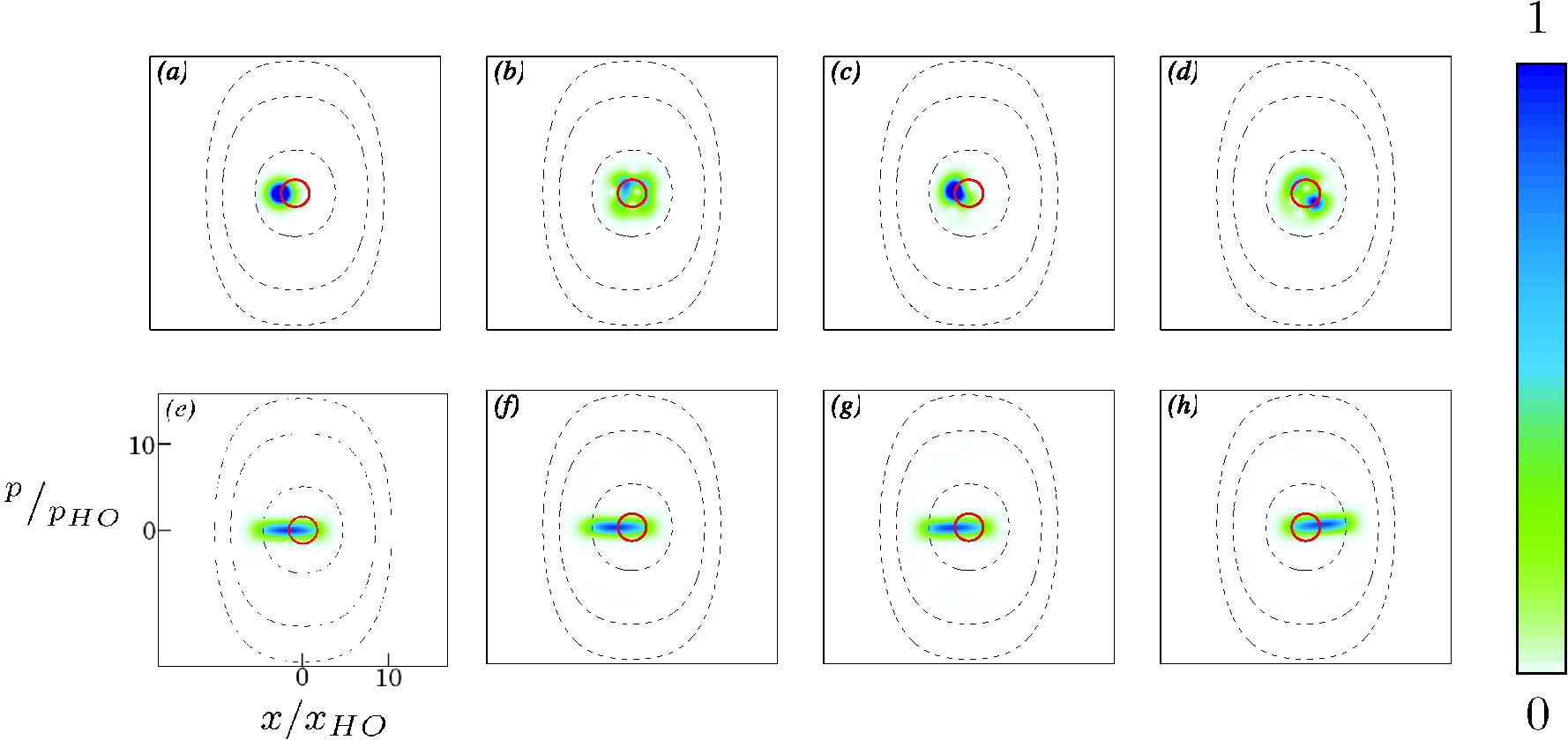}
\caption{(color online) Time evolution of the linear Schr\"odinger equation ($\gamma=0$, top panel), and of the GPE ($\gamma=115$, bottom panel) in the presence of an anharmonic (quartic) confining potential with parameters $\alpha=0.01$, and $\beta=0.04$. For each panel, the initial state is the corresponding (linear or nonlinear) coherent state, computed for $\alpha=0$. Husimi representations of the wavefunctions are given at times $\omega t/ (2\pi)=0$ (a) and (e), $3$ (b) and (f), $12$ (c) and (g), $40.5$ (d) and (h).}\label{fig5}
\end{figure*}

\begin{figure}
\includegraphics[width=1.0\linewidth,angle=0]{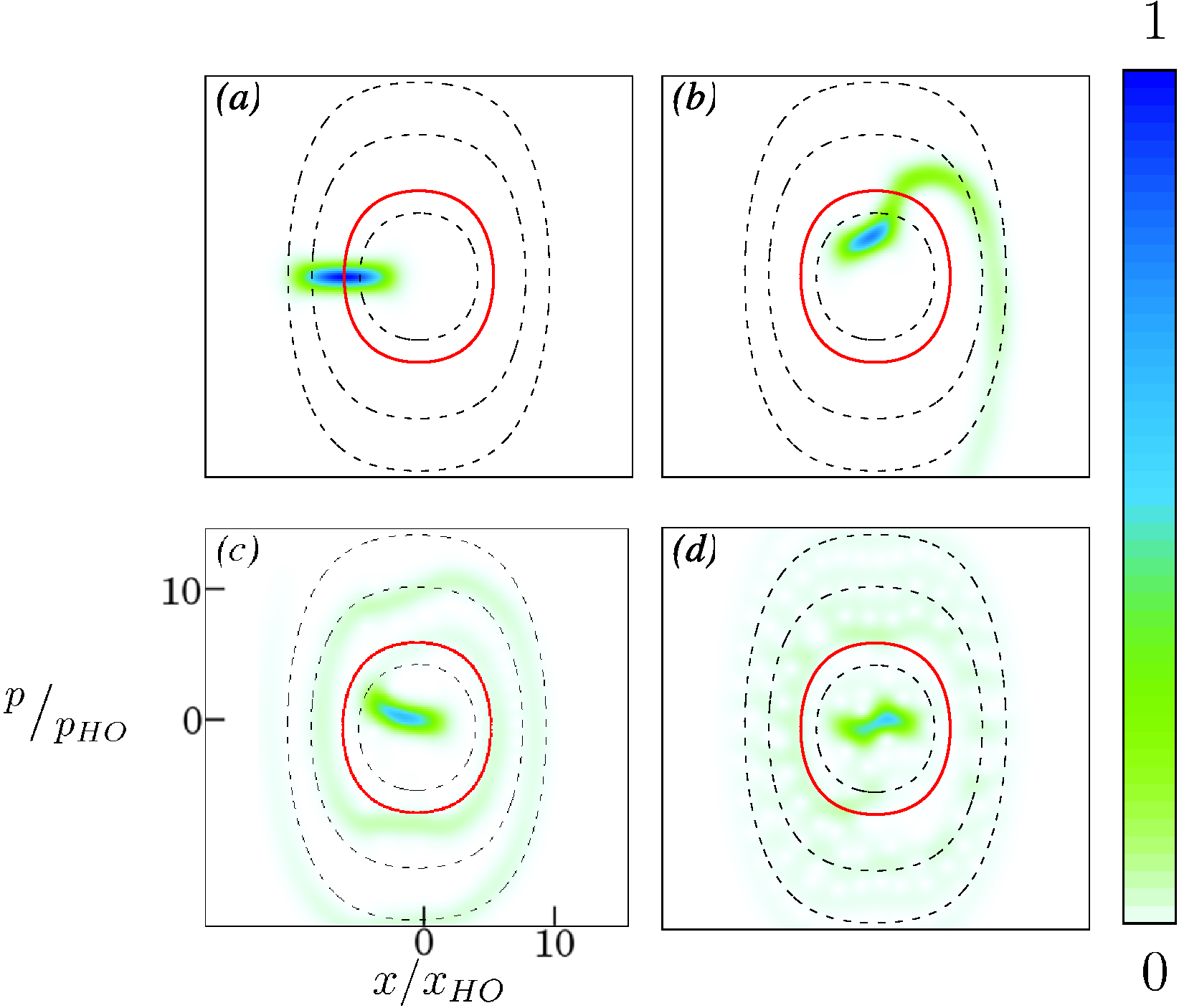}
\caption{(color online) Time evolution of the GPE in the presence of an anharmonic (quartic) confining potential, with parameters $\gamma=115$, $\alpha=0.01$, and $\beta=0.5$ . Husimi representation of the wavefunction at times $\omega t/ (2\pi)=0$ (a), $1.5$ (b), $4$ (c), $40$ (d).}\label{fig7}
\end{figure}

Figure~(\ref{fig5}) shows the time evolution for $\beta=0.04$. Before analyzing the results, it is useful to show the time evolution in the linear case. In the absence of nonlinear terms in the Schr\"odinger equation the time evolution is made of cycles of spreadings of the wave packet followed by a revival, i.e. after the spreading the packet comes back, to a good approximation, to its initial state, and the process starts again. This is indeed what is observed when $\gamma=0$ for an arbitrary value of $\beta$, see upper part of Fig.~(\ref{fig5}).

The motion of the corresponding coherent state in the presence of nonlinearities is quite different. For small values of $\beta$, such as the bottom panel of Fig.(\ref{fig5}), we observe that the nonlinear dynamics is more robust than the linear one. For such values of $\beta$ no spreading is observed. The packet keeps, to a good approximation, its initial shape during the time evolution, while the center follows the classical trajectory. Small amplitude dipole oscillations are observed, as well as a periodic motion of the tilting angle of the axis of the packet with respect to the $x$-axis. But roughly the packet (e.g. the condensate) preserves its coherence.

\begin{figure}
\includegraphics*[width=1.0\linewidth,angle=0]{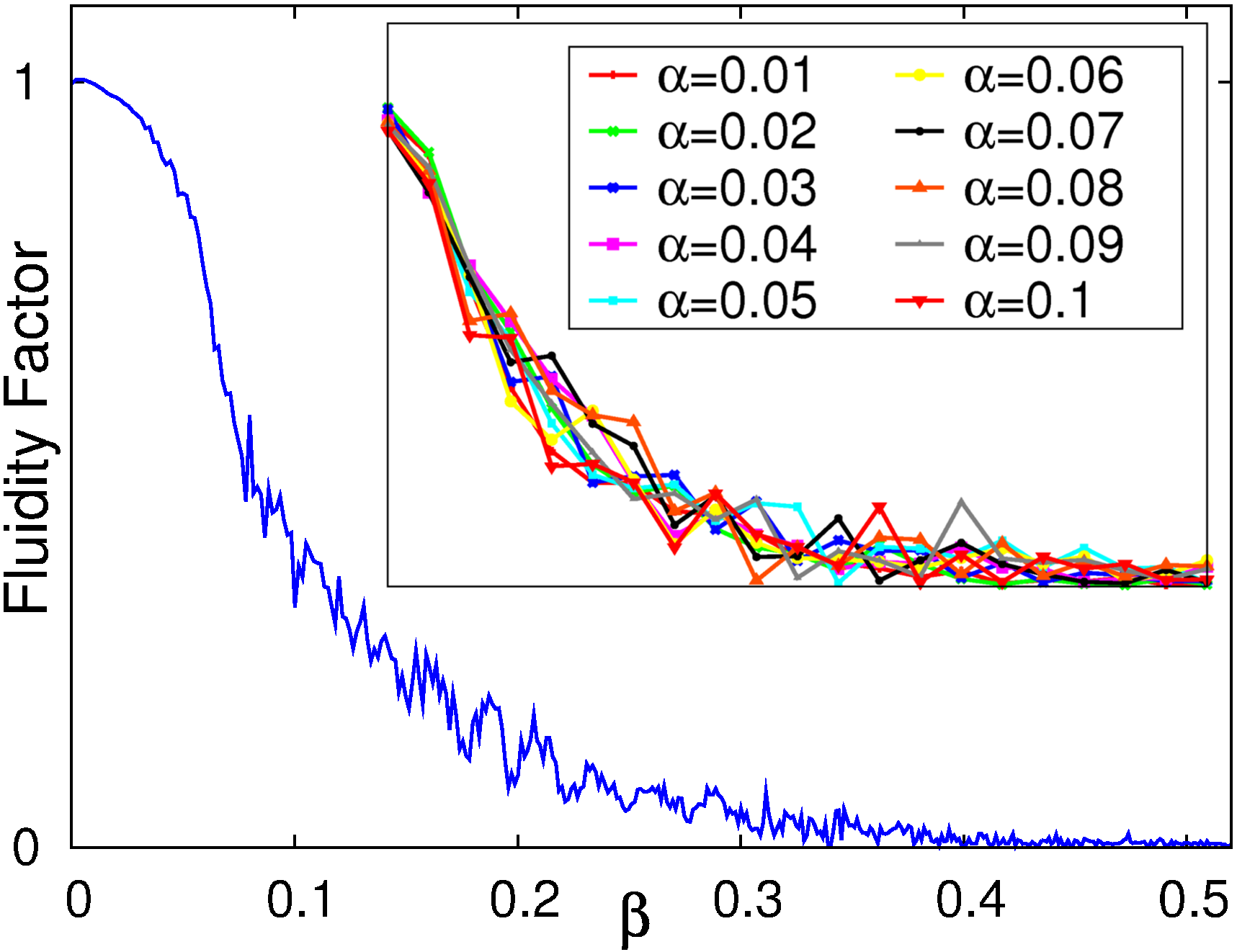}
\caption{(color online) The fluidity factor, defined by the ratio of the average amplitude of oscillation in a stationary regime and the initial amplitude $d$, versus $\beta = \alpha d^2$, for $\gamma=115$, and $\alpha=0.01$. The initial state is a shifted nonlinear coherent state. The inset shows the same plot for different values of $\alpha$, using for both axis the same scale as the main figure.}\label{fig6}
\end{figure}

Things change qualitatively as $\beta$ increases, as shown in Fig.~(\ref{fig7}). For larger initial amplitudes of the oscillation, at fixed $\alpha$, a strong deformation of the packet is observed during its time evolution. The packet does not preserve anymore its coherence. As it evolves, a filamentary structure develops from the packet and winds in the clockwise direction around it. This filament extends up to very high energies (see part (b) of the figure). By energy and mass conservation, the remaining packet has a smallersize and its center now occupies classical orbits of smaller energy, e.g., its amplitude of oscillation decreases. As time goes on, the winding filament compresses towards the packet. In this process, the different loops of the filament start to interfere. Finally, the reduced packet is completely damped at the bottom of the well, and coexists with a low density component which occupies a large fraction of the higher energy phase space, as shown in part (d) of Fig.~(\ref{fig7}). 

This is a remarkable process, that completely differs from what is known from the time evolution of the linear Schr\"odinger equation. Using the language of Bose-Einstein condensates, one can summarize it as follows (a similar effect is expected for, e.g., light motion in a nonlinear medium). In the presence of anharmonicities, the kinematic energy stored as center of mass motion of the condensate is not preserved, as for a harmonic potential. Instead, during the  dynamical evolution, one observes the emergence of two components. The initial packet is not totally destroyed. In the course of time, it looses part of its mass, and its amplitude of oscillation diminishes, to eventually be almost stopped at the bottom of the potential. The fraction of the condensate that leaves the packet occupies, in the course of time, high energy trajectories in an uncoherent way. One may speak of some sort of evaporative process, where the initial kinematic energy of the center of mass is transformed during the time evolution into uncoherent motion of high energy particles (evaporation), whereas the remaining fraction of the condensate cools down towards the bottom of the potential (damping effect).

At a given evolution time, the amplitude of oscillation of the remaining packet depends on $\beta$. To illustrate this point, we have computed, as a function of $\beta$, the time average of $2[ \langle x \rangle (t) / d]^2$, where $\langle x \rangle (t) = \int_{-\infty}^{\infty} x |\psi (x,t) |^2 dx$. The time average is computed for long times, starting from a time such that the evolution of $\langle x \rangle (t)$ looks stationary in time. This factor, that we call the fluidity factor, is equal to one at $\beta = 0$ (no damping of the wave packet) and equal to zero for a packet totally damped, almost at rest at the bottom of the potential. The result is represented in Fig.~\ref{fig6}. A strong decrease is observed as $\beta$ increases. For small values of $\beta$, there is no plateau where strictly no damping is observed. The fluctuations are due to the interactions between the low-density high-energy component with the main wave packet component. It may well be that if we further increase in time  the position of the time average window the fluidity factor globally decreases. That would mean that at very long times the packet is always fully damped. We cannot give for the moment a definite answer to this point.

We have also explored the dependence of this process on the different parameters. The inset of Fig.~\ref{fig6} shows the dependence of the fluidity factor on $\beta$ for packets that propagate in potentials with different values of $\alpha$. The superimposition of the curves shows that, on average, this quantity depends on $\alpha$ and $d$ only through $\beta=\alpha d^2$.

\section{Concluding remarks}

We have shown the existence of non-spreading states for the repulsive GPE, the so called nonlinear coherent states. They are defined as phase-space translations of the ground state of the nonlinear equation in presence of a harmonic confining potential. Due to the repulsive interaction, they are strongly elongated in the spatial direction. In the presence of a harmonic potential, the nonlinear coherent states do not vary their shape during the time evolution, their center simply follows a corresponding classical trajectory (of the linear problem). This means that the center of mass motion is decoupled from other modes of  the system. In particular, they are stable under shape deformations. We have computed the corresponding frequencies of oscillation for different nonlinearities. In the presence of a harmonic potential, the nonlinear coherent states thus preserve their coherence during the time evolution.

The physics is quite different when the nonlinear coherent states evolve in an anharmonic potential. We found that the time evolution now leads to a partial destruction of the initial packet (or of the condensate in BECs). During the time evolution, the system splits into two components. A fraction of the initial density leaves the packet, to occupy high energy phase space trajectories (evaporative process). The remaining fraction of the packet continues to oscillate around the bottom of the well but, by energy conservation, its amplitude now decreases (damping process). The anharmonicity of the potential thus induces a coupling between the dipole mode and other excitation modes. The initial center of mass kinematic energy is now partially transferred to a fraction of the particles, that leave the system, while the amplitude of the collective dipole motion of the remaining coherent component is damped. This process depends on the anharmonicity and on the initial amplitude through the parameter $\beta$, with a stronger damping for stronger values of $\beta$. 

In the presence of interactions, the revival phenomenon that occurs in linear quantum mechanics thus disappears and is replaced by a totally different mechanism. In the language of cold atom physics, the condensate is partially destroyed and damped when it evolves in an anharmonic confining potential.

Coherent transport and superfluidity are often tested by adding an external perturbation, like for instance the study of the damping of dipolar oscillations in BECs in the presence of an obstacle \cite{appl,hulet}. In 1D, the dissipative mechanism that breaks superfluidity is related to the emission of solitons. Also, loss of coherence and damping of collective excitations are predicted as temperature increases \cite{gzds}. Here, we have shown loss of coherence and dissipative effects in the absence of obstacles, simply induced by the presence of anharmonicities in the confining potential. 

Many interesting problems remain open, like a study of the motion of initial nonlinear packets in higher dimensional potentials, integrable or chaotic. The nature of the evaporative process described here should be further investigated using methods that go beyond the mean field approximation. Many studies already exist for the propagation of 1D packets in the presence of random potentials \cite{dima1, dima2, appl, appl2}. However, the present one-dimensional results show already the deep differences that exist between the linear and nonlinear cases in the presence of simple potentials.
Experimental tests of these differences are relatively easy, in particular in the cold atom context, by shifting a BEC with respect to an anharmonic potential.  In this article we have also explored the nonlinear dynamics of gaussian wave packets in both harmonic and anharmonic potentials, a problem that is relevant in optics experiments.

\section{Acknowledgements}

We thank M. Albert and N. Pavloff for fruitful discussions, and T. Paul for providing us a nonlinear Schr\"odinger equation program. This
work was supported by the ECOS-Sud grant A09E05.


\begin{thebibliography}{99}
\bibitem{com1} The same is true for the linear problem, where any eigenstate of the harmonic potential is a coherent state (with an increasing number of nodes), though the standard one is defined by the Gaussian ground state.
\end{thebibliography}
\end{document}